\documentclass{article}[12pt,psfig]
\usepackage{amsmath}
\usepackage{latexsym}
\usepackage{float}
\usepackage{amssymb}
\usepackage{epsfig}
\usepackage{graphicx}
\usepackage{cite}

\begin{document}

\title{Transport phenomena and microscopic structure in partially miscible binary fluids: A simulation
study of the symmetrical Lennard-Jones mixture}

\author{Subir K. Das, J\"urgen Horbach, and Kurt Binder\\
        \textit{Institut f\"ur Physik, Johannes Gutenberg-Universit\"{a}t} \\
        \textit {D-55099 Mainz, Staudinger Weg 7, Germany}}
\date{}

\maketitle

\renewcommand{\baselinestretch}{1.1}

\begin{abstract}
Static and dynamic structure factors and various transport coefficients are
computed for a Lennard-Jones model of a binary fluid (A,B) with a
symmetrical miscibility gap, varying both temperature and relative
concentration of the mixture. The model is first equilibrated by a
semi-grandcanonical Monte Carlo method, choosing the temperature
and chemical potential difference $\Delta \mu$ between the two
species as the given independent variables. Varying for $\Delta
\mu=0$ the temperature and particle number $N$ over a wide range,
the location of the coexistence curve in the thermodynamic limit is
estimated. Well-equilibrated configurations from these Monte Carlo
runs are used as initial states for microcanonical Molecular Dynamics
runs, in order to study the microscopic structure and the behavior of
transport coefficients as well as dynamic correlation functions along the
coexistence curve.  Dynamic structure factors $S_{\alpha \beta} (q,t)$
(and the corresponding static functions $S_{\alpha \beta} (q)$)
are recorded ($\alpha, \beta, \in$ A,B), $q$ being the wavenumber and
$t$ the time, as well as the mean square displacements of the particles
(to obtain the self-diffusion constants $D_{\rm A}$, $D_{\rm B}$) and
transport coefficients describing collective transport, such as the
interdiffusion constant and the shear viscosity. The minority species
is found to diffuse a bit faster than the majority species. Despite
the presence of strong concentration fluctuations in the system the
Stokes-Einstein relation is a reasonable approximation.
\end{abstract}
\newpage

\section{Introduction}

While the interplay between static structure and dynamical properties
of simple fluids is a problem intensively studied both by analytic
theory \cite{1,2,3,4,5} and simulations \cite{5,6,7,8,9,10}, and
also extensions to rather complex fluids such as polymers are
already given \cite{9,10,11,12,13,14}, somewhat less attention
has been paid to simple liquid mixtures: while static properties
of liquid-liquid miscibility have been analyzed in some detail
both by analytic theory (see e.g.~\cite{15}) and simulations (see
e.g.~\cite{5,16,17,18,19,20,21,22,23,24,25,26,27,28,29,30,31,32,33,34,35,36}),
dynamic properties of fluid mixtures have been investigated by computer
simulation only occasionally (e.g.~\cite{5,35,37,38,39,40,41,42}). In
most of these studies, the precise phase behavior of the used model is
not known, and also very small system sizes were used, and hence the
proper interpretation of the results is difficult.

The present work aims to contribute towards filling this gap by studying
both phase behavior and static structure in conjunction with dynamic
correlations and transport phenomena.  We address the simplest possible
case, a symmetric binary Lennard-Jones mixture with a miscibility
gap. Extensions of our approach to asymmetric Lennard-Jones mixtures and
models suitable for a realistic description of binary mixtures of liquid
metals (such as \cite{35}) will be given in future work. We note that a
specific asymmetric binary Lennard-Jones mixture with 80\% A and 20\%
B particles is a very popular model for the study of the dynamics of
glass-forming fluids \cite{43,44,45}. However, the static phase diagram
of that model system is not yet known.

The investigation of a symmetric mixture is of particular interest,
because one can clearly distinguish in this case long--ranged correlations
in the vicinity of the critical point from the different local ordering
of A- and B- particles that drives the unmixing transition well below
the critical temperature $T_{\rm c}$: At the critical point a mixture of 50\%
A- and 50\% B-particles is formed, i.e.~$x_{\rm B}=1-x_{\rm A}=N_{\rm
B}/(N_{\rm A}+N_{\rm B})=0.5$ ($N_{\rm A}$ and $N_{\rm B}$ being the
number of A and B particles in the simulation box, respectively), and
due to the symmetry of the model static A-A and B-B correlations are
identical in the latter case.  Well below $T_{\rm c}$ at coexistence one of
the species, say B, is the minority species and a B particle prefers a
larger number of neighboring B particles than in the one-phase region at
the same concentration of B particles. We shall discuss below how this
interplay of long-ranged correlations in the vicinity of $T_{\rm c}$
and local ordering far below $T_{\rm c}$ along the coexistence states
affects the dynamics of the mixture.

For a symmetric mixture the phase diagram and other static properties
are most easily computed with Monte Carlo (MC) methods using the
semi-grandcanonical ensemble rather than Molecular Dynamics (MD) methods.
This type of MC method has been used with great success to establish
binary alloy phase diagrams in the solid phase, both for lattice
\cite{46,47,48,49} and off-lattice \cite{50} models. It was also used to
study lattice models of polymers \cite{16,25,27,29}, and, in conjunction
with finite size scaling methods, to investigate the critical behavior
of symmetric binary Lennard-Jones fluids \cite{32}. It is also suitable
for asymmetric models of fluid mixtures \cite{17,35}. In this work we
combine semi-grandcanonical MC simulations with microcanonical MD runs
(i.e.~at constant energy $E$ and constant volume $V$).  The former
simulations yield well-equilibrated configurations at concentrations
$x_{\rm B}=1-x_{\rm A}$ and temperatures $T$ along the two-phase
coexistence curve of the system and thus, these configurations can be
used as initial states for the MD to study the structure and dynamics
at coexistence and also in the vicinity of coexistence by changing the
temperature or the composition.

The rest of the Paper is organized as follows. In the next section we
shall briefly comment on the model and the used simulation techniques.
A precise definition of the quantities that are computed is given
in Sec.~3. Then we present the results for the structure (Sec.~4)
and dynamics (Sec.~5) of our model that have been obtained from MD
runs. Finally, Sec.~5 presents our conclusions and gives an outlook on
further work.

\section{Model and simulation methods}

We consider a binary mixture of point-like particles interacting with 
Lennard-Jones potentials ($\alpha, \beta \in$ A,B),
\begin{equation} \label{eq1}
u(r_{ij}) = 4\varepsilon_{\alpha \beta} \; 
\left[ 
\left(\frac{\sigma_{\alpha \beta}}{r_{ij}} \right)^{12} - 
\left(\frac{\sigma_{\alpha \beta}}{r_{ij}} \right)^6 
\right], \quad r_{ij}=|\vec{r}_i-\vec{r}_j|,
\end{equation}
where $\{\vec{r}_i\}$ are the positions of the particles, and the
Lennard-Jones parameters $\varepsilon_{\alpha \beta}$ are chosen as
follows
\begin{equation} \label{eq2}
\sigma_{\rm AA}= \sigma_{\rm BB}=\sigma_{\rm AB}=\sigma ,\quad
\varepsilon_{\rm AA}=\varepsilon_{\rm BB}=\varepsilon, \quad 
\delta \equiv \varepsilon_{\rm AB}/\varepsilon .
\end{equation}
We measure all lengths in units of $\sigma \equiv 1$, and we also choose
Boltzmann's constant $k_B \equiv 1$. The temperature $T$ will be measured
in units of $\varepsilon \equiv1$, and (in the MD part) the masses of the
two particle species are also chosen the same, $m_{\rm A}=m_{\rm B}=1$.
The potential is truncated and shifted at $r_{ij}=2.5 \, \sigma$.

The same model has been studied earlier by Wilding \cite{32} (note that
in this work the potential was also truncated at $r_{ij}=2.5 \, \sigma$
but not shifted, and thus minor differences in the properties between
our model and that of Ref.~\cite{32} must be expected). In the latter
work, also the density of the binary fluid was varied over a wide range,
$0 \leq \rho \leq 0.7$, and it was found for $\delta =0.7$, that the
$\lambda$-line of critical points for fluid-fluid phase separation ends
at a critical end point at a density of about $\rho \sigma^3 \approx 0.6$
at the coexistence curve of liquid-gas phase separation. At sufficiently
smaller values of $\delta$ than 0.7 the $\lambda$ line moves towards
the liquid-gas critical point where it merges into a tricritical point
and one may expect that this happens at densities below $\rho \sigma^3
\approx 0.6$~\cite{32}.  In the present work, we are not interested
in the liquid-gas transition of the model at all, and hence work at a
much higher density, namely $\rho=1.0$. For this density we stay in the
fluid phase for the whole temperature range of interest, i.e.~$1 \leq T
\leq 1.8$.  Choosing $\delta =0.5$, the critical temperature of unmixing
is $T_c\approx1.638 \pm 0.005$ at this density, see Fig.~\ref{fig2} below.

\begin{figure}
\psfig{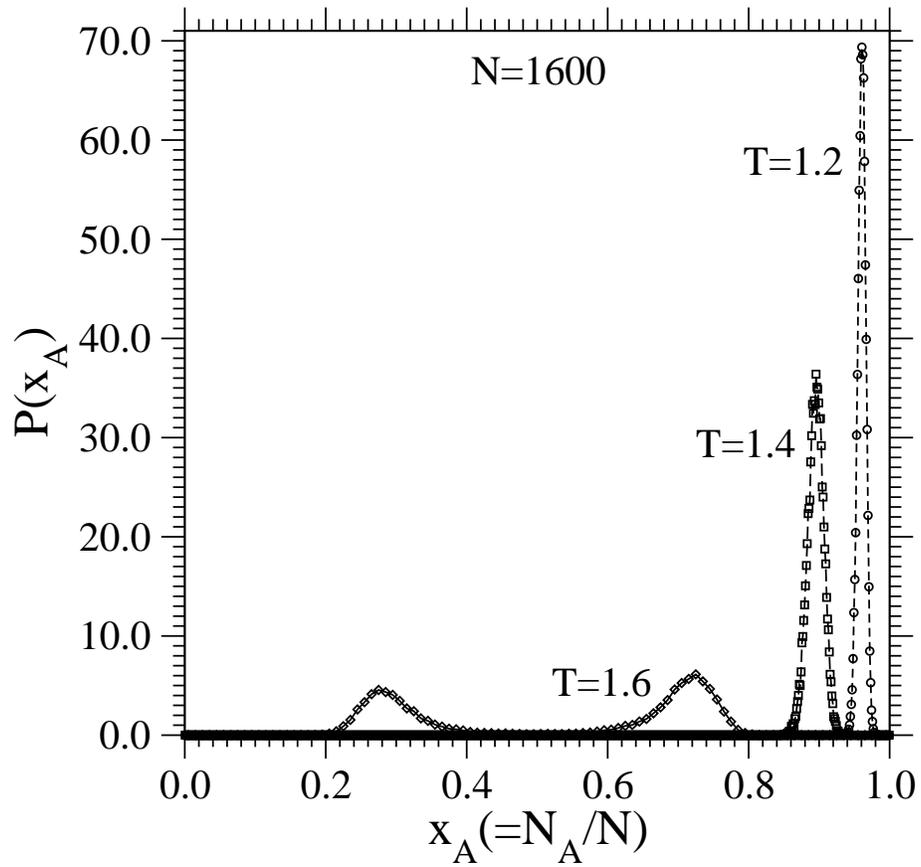}
\caption{\label{fig1}
Distribution function $P(x_{\rm A})$ for the
relative concentration $x_{\rm A}=N_{\rm A}/N$ of A particles for $N=1600$
particles at density $\rho=1$ and the three indicated temperatures.}
\end{figure}

The equilibration of the model is done as follows.  We start from random
configurations with an equal number of A and B particles. The particles
are then displaced by means of a standard Monte Carlo scheme in the NVT
ensemble, i.e.~at constant total particle number $N$, constant volume
$V$, and constant temperature $T$, rejecting or accepting a trial move
according to the standard Metropolis criterion.  Thereby, the length
of a trial displacement is a randomly chosen number in the interval
$[-\sigma/20,\sigma/20]$. After 100000 MC steps the semi-grandcanonical
MC is switched on: At the end of each displacement step an attempted
identity switch (B $\longrightarrow$ A or A $\longrightarrow$ B) of
$N/10$ randomly chosen particles is made, which is again accepted or
rejected according to a standard Metropolis criterion, where both the
energy change $\Delta E$ and the change of the chemical potential
$\pm \Delta \mu$ needs to be taken into account in the Boltzmann
factor \cite{7,9,10,50,51}. Each configuration after this Monte Carlo
move (whether the change was accepted or not) is counted as a state
for the Monte Carlo averaging in the semi-grandcanonical ensemble,
which is defined by the independent variables $N$, $V$, $T$, and the
chemical potential differences $\Delta \mu$. Note that the relative
concentration $x_{\rm B} = N_{\rm B}/(N_{\rm A}+N_{\rm B})=N_{\rm B}/N$
is a fluctuating quantity and its average value is an output rather
than an input to the simulation. Only for the choice $\Delta \mu=0$ the
symmetry of the model between A and B dictates that for $T \geq T_c$ we
must have $\langle x_{\rm B} \rangle=1/2$, while for $T<T_c$ we obtain
for $\Delta \mu=0$ states at the two-phase coexistence curve.  This is
illustrated in Fig.~\ref{fig1}, where the concentration distribution
function $P(x_{\rm A})$ is plotted for three temperatures. Note that we
have done 5 independent runs, each with a length of 400000 MC steps in the
semi-grandcanonical ensemble, to calculate the distribution functions in
Fig.~\ref{fig1} whereby we started the averaging after 100000 MC steps in
each run. In principle, for a finite system there is no symmetry breaking
possible in full thermal equilibrium, and hence one should observe
always two peaks, one at $x_{\rm A}^{\textrm{coex}(2)}$ and the other
one at $x_{\rm A}^{\textrm{coex}(1)}=1-x_{\rm A}^{\textrm{coex}(2)}$,
corresponding to the two branches of the two-phase coexistence curve. In
practice, these two branches are separated at low temperatures by a
huge free energy barrier in phase space. Therefore, as can be seen
in Fig.~\ref{fig1}, at $T=1.2$ and $T=1.4$ only the A-rich phase is
observed, and no transition to the B-rich side of the coexistence curve
has happened yet at all. Only for $T=1.6$, i.e.~rather close to $T_c$
(which in Fig.~\ref{fig1} is evident from the large width of the peaks),
transitions back and forth between the two coexisting phases have occurred
(although there were not enough of those transitions to make $P(x_{\rm
A})$ strictly symmetric around $x_{\rm A}^{\textrm{crit}} = 1/2$).

This ``ergodicity breaking'' due to finite observation time in the
context of first-order phase transitions is a well-known phenomenon
\cite{51} and does not hamper our analysis at all. From data such as
shown in Fig.~\ref{fig1} we hence obtain, when we normalize the part
of $P(x_{\rm A})$ for $x_{\rm A} > 1/2$ to $\int P (x_{\rm A}) d x_{\rm
A}=1$, both estimates for the location of the coexistence curve and the
``concentration susceptibility'' $\chi$,
\begin{equation} \label{eq3}
x_{\rm A}^{\textrm{coex} (2)} = 
\int\limits_{1/2}^1 \, x_{\rm A} P (x_{\rm A}) d x_{\rm A}
\quad , \quad 
k_BT \chi= N \left( 
\int\limits_{1/2}^1 x_{\rm A}^2 P(x_{\rm A}) d x_{\rm A}- 
[x_{\rm A}^{\textrm{coex} (2)}]^2 \right) .
\end{equation}
In order to investigate the dynamics of well-equilibrated states
precisely at the coexistence curve, we store a number of $n=10$
independent configurations at each of the 5 independent runs which
have a number $N_{\rm A}$ of A particles such that the resulting ratio
$x_{\rm A}$ is as close to $x_{\rm A}^{\textrm{coex}(2)}$ as possible
(note that $x_{\rm A}$ is ``quantized'', because $N_A$, $N$ are integers,
but for large $N$ the discreteness of $x_A$, which is $1/N$, is smaller
than the statistical error with which $x_{\rm A}^{\textrm{coex} (2)}$
can be estimated, and therefore this effect does not matter). With
these configurations we can realize microcanonical MD runs right at the
coexistence curve. Note that we have used in this work a standard MD
code applying the velocity Verlet algorithm with a time step $\delta t =
0.01$ \{in units of the time $t_0=(m \sigma^2 / 48 \varepsilon)^{1/2}$\}.

When we work in the one phase region we just heat up the configurations
that we have gotten from the semi-grandcanonical Monte-Carlo
at coexistence and we equilibrate the system by MD at the desired
temperature $T_{{\rm f}}$.  The temperature is then kept constant by
coupling the system to a stochastic heat bath, i.e.~every 50 MD steps
the velocities are replaced by new ones that are chosen randomly from
a Boltzmann distribution corresponding to the temperature $T_{{\rm f}}$.

We have described these simple and almost straightforward
considerations~\cite{54} in such detail here, because the literature
is still full of examples where the authors are not aware of the
pitfalls presented to equilibration by the presence of slowly relaxing
variables. In the present problem, the conservation of the concentration
(in the ensemble where both $N_{\rm A}$ and $N_{\rm B}$ are held fixed)
together with the presence of a large correlation length of concentration
fluctuations (which for $N \rightarrow \infty$ diverges at the critical
point and causes the critical slowing down \cite{53}) may lead to a very
slow relaxation. Hydrodynamic slowing down is eliminated by the choice
of an ensemble where the slow variable is not conserved, in our case the
semi-grandcanonical ensemble. Critical slowing down cannot be avoided
(apart from the special case of Ising models on lattices, for which
cluster algorithms have been put forward \cite{55,56} that eliminate also
critical slowing down to a large extent). In the present work, we hence
do not attempt to study the immediate vicinity of the critical point.

Fig.~\ref{fig2} shows the coexistence curve as estimated from our
semi-grandcanonical method. One can see clearly that for $T \leq 1.4$
finite size effects on the coexistence curve are quite negligible, if $N
\geq 400$. For $T=1.5$, data for $N=400$ overestimate the width of the two
phase region very slightly, while data for $N=$ 800, 1600 and 3200 agree
within statistical errors. For $T=1.6$, however, data for both $N=400$
and $N=800$ overestimate the width of the coexistence curve slightly,
while the data for both $N=1600$ and $N=3200$ still agree.  For this
reason, we have chosen $N=1600$ as the standard particle number, for
which all analysis of static and dynamic correlations have been made
(Secs.~3, 4). We have estimated the critical temperature $T_c$ from
power law fits of the form
\begin{equation}
  f(x_{\rm B})= 1/2 \pm x_{\rm B}^{\textrm{coex}} = \widehat{B} (1-T/T_c)^\beta
  \label{eq_pow}
\end{equation}
with the critical exponent $\beta=0.33$ which corresponds to the
universality class of the three-dimensional Ising model. $\widehat{B}$
and $T_c$ have been used as adjustable parameters. For $N=400$ the fit
yields $T_c \approx 1.666 \pm 0.005$, while for all larger values of
$N$ the result is $T_c\approx 1.638 \pm 0.005$ (see Fig.~\ref{fig2}).
We emphasize at this point, however, that we deliberately do not
address the problem of analyzing the critical behavior of transport
coefficients in our study: for such a purpose one would need to work
with system sizes $N$ that are several orders of magnitude larger, and
extremely accurate data in the temperature range $1.6 \leq T \leq 1.7$
would be needed. This interesting problem would require substantially
more computer resources than were available to us. In addition, we would
have to locate $T_c$ much more accurately with a finite size scaling
analysis \cite{9,16,25,27,29,32,50,51}. Even though we do not study
critical behavior, our choice of $N$ is much larger than most of the
choices made in previous work on the dynamics of mixtures (e.g.~Vogelsang
and Hoheisel \cite{39} work with $N=108$ and $N=256$ particles, and most
data of Asta {\it et al.}~\cite{35} refer to $N=500$ particles).

\section{Static and dynamic quantities}

We now proceed to define the quantities that are computed in our
simulations. With respect to static quantities, we have monitored the
standard pairwise radial distribution functions $g_{\alpha \beta} (r)$
between the different pairs $\{\alpha, \beta \in ($A,B$)\}$ of particles
\cite{1,2,3,4,5,6,7,8},
\begin{equation}
 g_{\alpha \beta}(r) = \frac{N}{\rho N_{\alpha} N_{\beta}}
              \left< \sum_{i=1}^{N_{\alpha}} 
              \sum_{j=1}^{N_{\beta}} {}^{\prime}
               \; \delta (r - |\vec{r}_i (0) - \vec{r}_j (t) | )
               \right> \quad .
\end{equation}
The prime in the second sum means that $i=j$ has to be left out if
$\alpha=\beta$.

In terms of the $g_{\alpha \beta}(r)$, partial structure functions
$S_{\alpha \beta} (q)$ are defined as follows~\cite{2}:
\begin{equation} \label{eq4}
S_{\alpha \beta} (q) = x_{\alpha} \delta_{\alpha \beta} + 
x_{\alpha} x_{\beta} \rho \int\limits_0^\infty g_{\alpha
\beta} (r) \,\, \frac{\sin (qr)} {q r} \,\, 4 \pi r^2 d r.
\end{equation}
with $x_{\alpha}=N_{\alpha}/N$ ($\alpha \in$ A, B). In order to compute
the partial structure factors we used the formula
\begin{equation}
S_{\alpha \beta}(\vec{q}) = \frac{1}{N} 
\sum_{k=1}^{N_{\alpha}} \sum_{l=1}^{N_{\beta}} 
\left< \exp (i \vec{q} \cdot \vec{r}_{kl}) \right> \quad ,
\end{equation}
and performed the angular integration numerically to obtain 
$S_{\alpha \beta}(q)$.

It turns out that it is more useful to form the following linear
combinations of the radial distribution functions $g_{\alpha \beta}$
and the partial structure factors $S_{\alpha \beta}$ which describe
correlations in number density ($g_{nn}$ and $S_{nn}$), concentration
correlations ($g_{cc}$ and $S_{cc}$), and cross correlations
between number density and concentration ($g_{nc}$ and $S_{nc}$)
\cite{2,57,58}. The real space functions are given by
\begin{eqnarray}
g_{nn}(r) & = & 
x_{\rm A}^2 g_{\rm AA}(r) + 2 x_{\rm A} x_{\rm B} g_{\rm AB} (q) + 
  x_{\rm B}^2 g_{\rm BB}(r) \quad , \label{eq5a} \\
g_{cc}(r) & = & 
x_{\rm A}^2 x_{\rm B}^2 \left[ g_{\rm AA}(r) + g_{\rm BB}(r) 
  - 2 g_{\rm AB}(r) \right]
\quad , \label{eq6a} \\
g_{nc}(r) & = & 
x_{\rm A} x_{\rm B} \left[ x_{\rm A} g_{\rm AA}(r) - x_{\rm B} g_{\rm BB}(r) 
   + (x_{\rm B}-x_{\rm A}) g_{\rm AB}(r) \right]
\quad , \label{eq7a}
\end{eqnarray}
and the functions in reciprocal space by
\begin{eqnarray}
S_{nn}(q) & = & S_{\rm AA} (q) + 2 S_{\rm AB} (q) + 
                S_{\rm BB} (q) \quad , \label{eq5} \\
S_{cc}(q) & = & 
x_{\rm B}^2 S_{\rm AA}(q) + x_{\rm A}^2 S_{\rm BB}(q) - 
2 x_{\rm A} x_{\rm B} S_{\rm AB}(q) 
\quad , \label{eq6} \\
S_{nc}(q) & = & 
x_{\rm B} S_{\rm AA}(q) - x_{\rm A} S_{\rm BB}(q) 
+ (x_{\rm B}-x_{\rm A}) S_{\rm AB}(q) 
\quad . \label{eq7}
\end{eqnarray}
We will see below that strong concentration fluctuations are observed
if one approaches the critical point of unmixing (and this happens not
only in the immediate vicinity of the critical point). One can try to
describe $S_{cc}(q)$ at small $q$ by the Ornstein-Zernike form,
\begin{equation}
S_{cc}(q) = \frac{k_B T \chi}{1 + \xi^2 q^2} \quad , \label{eq8a}
\end{equation}
where the ``susceptibility'' $\chi$ is given by Eq.~(\ref{eq3}) and $\xi$
has the meaning of a static correlation length.

Turning to dynamic quantities, the most straightforward quantity to
consider is the mean square displacement of tagged particles
\begin{equation} \label{eq8}
g_{\rm A} (t)= 
\left< [\vec{r}_{i, {\rm A}} (0) - \vec{r}_{i, {\rm A}} (t)]^2
\right>, \quad g_{\rm B} (t) = \left< [\vec{r}_{j, {\rm B}} (0) -
\vec{r}_{j, {\rm B}} (t)]^2 \right> \quad,
\end{equation}
where it is understood that the average $\langle \cdots \rangle $ includes
an average over all particles of type A or B, respectively. Another
quantity which also monitors the motion of individual particles are
the incoherent intermediate structure functions $F^{\rm (A)}_s (q,t)$
and $F^{\rm (B)}_s (q,t)$,
\begin{eqnarray}
F^{\rm (A)}_s(q,t) & = & 
\frac{1}{{N}_{\rm A}} \, \sum\limits_{i \, \in  {\rm A}}
\langle \exp \{-i \, \vec{q} \cdot [\vec{r}_i(0) -\vec{r}_i
(t)]\}\rangle \quad , \label{eq9} \\
F_s^{\rm (B)} (q,t) & = & 
\frac{1} {N_{\rm B}} \, \sum\limits_{j \, \in {\rm B}} \,
\langle \exp \{-i \vec{q} \cdot [\vec{r}_j (0) - \vec{r}_j (t)]\}
\rangle  \quad , \label{eq10}
\end{eqnarray}
where the sum runs over all A particles \{Eq.~(\ref{eq9})\} or B particles
\{Eq.~(\ref{eq10})\}, respectively. Note that due to the isotropy of the
fluid, Eqs.~(\ref{eq9}), (\ref{eq10}) cannot depend on the direction of
the scattering vector $\vec{q}$.  When the displacements of the particles
are Gaussian distributed, Eqs.~(\ref{eq9}), (\ref{eq10}) reduce to
\begin{equation} \label{eq11}
F_s^{\rm (A)} (q,t) = 
\exp \left[- \frac{1}{6} q^2 g_{\rm A} (t) \right], \quad
F^{\rm (B)}_s (q,t) = 
\exp \left[-\frac{1} {6} q^2 g_{\rm B} (t) \right].
\end{equation}
Noting further the Einstein relations 
\begin{equation} \label{eq12}
g_{\rm A}(t)= 6 \, D_{\rm A} t, \quad g_{\rm B}(t)= 6 D_{\rm B} t , 
\quad t \rightarrow \infty , 
\end{equation} 
where $D_{\rm A}$ and $D_{\rm B}$ are self-diffusion constants of the
particles, we expect that the asymptotic decay of $F^{\rm (A)}_s (q,t)$
and $F^{\rm (B)}_s (q,t)$ can be described by a simple exponential
variation with time,
\begin{equation} \label{eq13}
F_s^{\rm (A)} (q,t) \propto \exp [- t/\tau_{\rm A}(q)] \, \, , \quad
F_s^{\rm (B)} (q,t) \propto \exp [-t/\tau_{\rm B}(q)] \,\, ,
\end{equation}
and Eqs.~(\ref{eq11})-(\ref{eq13}) then imply
\begin{equation} \label{eq14}
\lim\limits_{q \rightarrow 0} [\tau_{\rm A} (q) q^2]^{-1} =
D_{\rm A}, \quad 
\lim\limits_{q \rightarrow 0} [\tau_{\rm B}(q) q^2]^{-1} = 
D_{\rm B} \quad .
\end{equation}

In addition, we introduce partial coherent intermediate structure
functions $S_{\alpha \beta} (q,t) $ defined as $[\alpha, \beta \in $
(A,B)]
\begin{equation} \label{eq15}
S_{\alpha \beta} (q,t) =
\frac{1}{N} \,\, \sum\limits_{{i=1}\atop{i
ß \, \in \,  \alpha}}^{N_{\alpha}} \,\, \sum\limits_{{j=1}\atop{j
\, \in \, \beta}}^{N_\beta} \,\left< \exp \{-i \vec{q} \cdot
[\vec{r}_i(0) - \vec{r}_j (t)] \} \right> \quad ,
\end{equation}
and from these functions we can define number density and concentration
correlation as well as cross correlation functions $\{ S_{nn} (q,t),
S_{cc} (q,t)$ and $S_{nc} (q,t)\}$ by forming exactly the same type
of linear combinations as written in Eqs.~(\ref{eq5})-(\ref{eq7}), but
with the $S_{\alpha \beta}(q,t)$ rather than their static counterpart
$S_{\alpha \beta} (q)$.  In order to compare the time correlation
functions $S_{\alpha \beta}(q,t)$ for different values of $q$ it is
convenient to consider the normalized functions
\begin{equation} \label{eq15a}
  F_{\alpha \beta}(q) = 
  \frac{ S_{\alpha \beta} (q,t) }{ S_{ \alpha \beta } (q) } \quad .
\end{equation}
The normalized functions $F_{nn}(q,t)$, $F_{cc}(q,t)$, and $F_{nc}(q,t)$
are defined in a similar way by normalizing $S_{nn}(q,t)$, $S_{cc}(q,t)$
and $S_{nc} (q,t)$ by the corresponding static correlation function.

Further quantities of interest are the instantaneous values of the
components of the pressure tensor ($x,y,z$ denote the Cartesian
components, $\vec{v}_i$ is the velocity of the $i$'th particle)
\begin{equation} \label{eq16}
\sigma_{xy} (t) = \sum\limits_{i=1}^N \left[ m_i v_{ix} v_{iy} +
\frac{1}{2} \sum\limits_{j(\neq i)} \, |x_i-x_j | F_y (|\vec{r}_i
-\vec{r}_j|) \right] \quad ,
\end{equation}
$\vec{F}$ being the force acting between particles $i,j$. From the
autocorrelation function of $\sigma_{xy} (t)$ one gets the shear
viscosity $\eta$ of the fluid by a Green-Kubo formula \cite{1,2},
\begin{equation} \label{eq17}
\eta=\frac{1}{Vk_BT} \, \int\limits_0^\infty d t \; \langle \sigma
_{xy} (0) \sigma _{xy} (t) \rangle \,\, .
\end{equation}

Finally we mention the Green-Kubo formula for the interdiffusion
constant (note that we have chosen the masses of all particles
equal to each other here) \cite{2}
\begin{equation} \label{eq18}
D_{\textrm{int}} =
\frac{1}{N S_{cc}(0)} \,\,
\int\limits_0^\infty dt \; \langle J_x^{\rm int} (0) \, J_x ^{\rm int}
(t) \rangle \quad ,
\end{equation}
where the current $\vec{J}^{\rm int} (t)$ is defined as follows:
\begin{equation} \label{eq19}
\vec{J}^{\rm int} (t)= 
x_{\rm B} \sum\limits_{i=1}^{N_{\rm A}} \, \vec{v}_i(t) -
x_{\rm A} \sum\limits_{i=1}^{N_{\rm B}} \, \vec{v}_i(t)
\quad .
\end{equation}
One can easily calculate $D_{\textrm{int}}$ for an ideal mixture [such
a mixture is formed if the particles are labeled but are otherwise
identical and thus the entropy of mixing is equal to $-Nk_B (x_{\rm A}
{\rm log} \, x_{\rm A} + x_{\rm B} {\rm log} \, x_{\rm B})$].  In this case
cross correlations in the autocorrelation function for $J_x^{\rm int}$ in
Eq.~(\ref{eq18}) vanish and the interdiffusion constant can be expressed
by the self-diffusion constants~\cite{2},
\begin{equation}
  D_{\textrm{int}} = x_{\rm B} D_{\rm A} + x_{\rm A} D_{\rm B} \ .
  \label{app_dint}
\end{equation}
It has been shown by MD simulations that Eq.~(\ref{app_dint}) is a good
approximation for mixtures of Lennard-Jones fluids in the one-phase
region~\cite{38,59}. However, for ideal mixtures on a rigid lattice
(where A, B-atoms hop with jump rates $\sigma_{\rm A}$, $\sigma_{\rm B}$
to vacant sites, assuming that a small number of vacancies is present)
Eq.~(\ref{app_dint}) does not hold~\cite{kehr}.

\section{Static properties of the symmetrical binary fluid mixture}

In this section we analyze the structure of the symmetric Lennard-Jones 
mixture. Results are shown along the coexistence line and along the different
paths in the one-phase region that are indicated in Fig.~\ref{fig2}. 

In Fig.~\ref{fig3} we present the pair correlation functions $g_{\alpha
\beta} (r)$ for the three states at temperatures $T=1.2$, 1.4, and 1.6 at
the coexistence curve (cf.~Fig.~\ref{fig2}). These data show a typical
normal fluid behavior in all cases, as expected.  Although it would be
hard to recognize from these partial radial distribution functions that
one approaches the critical point of unmixing in a binary fluid, the
comparison of the different correlations shows nontrivial features. At
each temperature, the first peak in $g_{\rm AB}(r)$ is at a slightly
smaller distance than in $g_{\rm AA}(r)$ and $g_{\rm BB}(r)$.  Moreover,
the second peak in $g_{\rm BB}(r)$ is shifted towards smaller distances
compared to the other two functions and this effect is more pronounced the
lower the temperature is, i.e.~the farther one is away from the critical
point. The explanation of this feature is simple: The B particles are
the minority species at the considered state points and thus, at small
concentrations $x_{\rm B}$ it is very likely that one finds an A particle
between two next-nearest B particles forming a B-A-B sequence. Since
the distance between nearest A-B neighbors is smaller than that between
nearest A-A and B-B neighbors one may expect that a large amount of such
B-A-B sequences, where the distance between next-nearest B particles is
smaller than in a B-B-B sequence, leads to the shift in the second peak
of $g_{\rm BB}(r)$. Certainly, the difference between $g_{\rm AA}$ and
$g_{\rm BB}$ becomes less pronounced the closer one approaches $x_{\rm
B}=0.5$ (and thus the critical point) where, due to the symmetry of our
model, $g_{\rm AA}$ and $g_{\rm BB}$ are identical.

The temperature dependence of $g_{\rm AA}(r)$ and $g_{\rm BB}(r)$
in the one-phase region is shown in Fig.~\ref{fig4} for the constant
concentration $x_{\rm B}=0.10375$ in the temperature range $1.4 \le T
\le 1.8$ ($T=1.4$ corresponds to a state on the coexistence curve). We
see that the temperature dependence is very weak for $g_{\rm AA}(r)$
whereas we observe a significant increase of the amplitude of the
first peak in $g_{\rm BB}(r)$ decreasing the temperature towards the
coexistence line.  The latter effect is even more pronounced in the
coordination number function $z_{\alpha \alpha}(r)$ that is shown in
the insets of Fig.~\ref{fig4}. This function is defined by
\begin{equation}
  z_{\alpha \alpha}(r) = 
  x_{\alpha} \int_0^r dr^{\prime} \; 
  4 \pi r^{\prime \, 2} g_{\alpha \alpha}(r^{\prime}) 
  \quad  \alpha \in {\rm [A,B]}
\end{equation}
which gives the number of particles of type $\alpha$ surrounding
a particle of type $\alpha$ within a distance $r^{\prime} < r$. In
the considered temperature range $z_{\rm BB}(r)$ changes at $r=1.4$
(i.e.~around the location of the first minimum) from $z_{\rm BB}=1.37$
for $T=1.8$ to $z_{\rm BB}=1.55$ for $T=1.4$.  By a closer inspection
of $z_{\rm AA}(r)$ one finds that the changes in this quantity are of
the same order. Thus there are only minor changes in the local order if
one approaches the coexistence line from above.

We now proceed to discuss the behavior of the structure factors
$S_{nn}(q)$, $S_{cc}(q)$, and $S_{nc}(q)$ \{Eqs.~(\ref{eq5})-(\ref{eq7})\}
along the coexistence line and for the latter two quantities also the
corresponding functions in real space \{Eqs.~(\ref{eq6a})-(\ref{eq7a})\},
see Fig.~\ref{fig5}. The number density structure factor $S_{nn}(q)$
behaves exactly as expected for any dense simple liquid, and in particular
$S_{nn}(q \rightarrow 0)$ is very small, as expected for liquids that
are almost incompressible. The height of the first peak is around 3, as
is typical for fluids at temperatures somewhat higher than the melting
temperature, and there is a pronounced second peak, due to the rather
regular close packing of atoms in a dense fluid.

In contrast, $S_{cc} (q)$ at $T=1.2$ is very small and structureless
throughout: in symmetrical binary mixtures, there is very little coupling
between density and concentration fluctuations. For $T=1.4$, we see
already some enhancement of $S_{cc}(q)$ at small $q$, reflecting the
growing concentration fluctuations. For $T=1.6$, at small $q$ a rather
dramatic growth of $S_{cc}(q \rightarrow 0)$ signals the proximity
of the critical point. For small $q$ we have fitted the data by the
Ornstein-Zernike form Eq.~(\ref{eq8a}) (bold lines in Fig.~\ref{fig5}b),
where we have used the correlation length $\xi$ as an adjustable
parameter and we have estimated the susceptibility $\chi$ by means of
Eq.~(\ref{eq3}) from the distribution function $P(x_{\rm A})$ (the values
of $S_{cc}(0)=k_BT \chi$ are shown as crosses in Fig.~\ref{fig5}b). From
the fits we recognize that the data for small $q$ can be described by
Eq.~(\ref{eq8a}) within the statistical errors (note that there is only
one adjustable parameter, the correlation length $\xi$).  From the fits
we find a growing correlation length $\xi$ with increasing temperature
towards $T_c$.  This behavior can be also inferred from $g_{cc}(r)$
(see inset of Fig.~\ref{fig5}b) at $T=1.6$ which has a rather pronounced
tail for $r>2$. Also remarkable in $g_{cc}(r)$ is the negative peak
around $r\approx 0.9$ which stems from the fact that these distances
are avoided by nearest A-A and B-B neighbors but are typical for nearest
A-B neighbors.

Particularly interesting is the structure factor $S_{nc}(q)$. In a
mixture with $\Delta \mu=0$ at $T>T_c$, where then $x_{\rm A}=x_{\rm
B}=1/2$, we expect that $S_{nc}(q) \equiv 0$ due to the perfect symmetry
between A and B. Along the coexistence curve, the spontaneous symmetry
breaking between A and B invalidates this argument, and hence a nonzero
$S_{nc}(q)$ is possible. As we see in Fig.~\ref{fig5}c this is obviously
the case. The origin of the oscillations in $S_{nc}(q)$ can be most easily
understood by means of its counterpart in real space, $g_{nc}(r)$ (see
inset of Fig.~\ref{fig5}c). The most pronounced feature in the latter
quantity is a peak with negative intensity around $r_1\approx 0.9$.
This feature is again due to the fact that the first peak in $g_{\rm
AB}(r)$ is slightly shifted to smaller distances compared to the first
peak in $g_{\rm AA}$ and in $g_{\rm BB}$ such that the intensity of the
latter two functions in the interval $0.85 < r < 0.9$ is essentially zero
whereas $g_{\rm AB}$ is rapidly increasing for the latter distances.
This leads to the occurrence of negative correlations in $g_{nc}$ and
thus to the peak around $r_1$. We emphasize that the scale for $S_{nc}(q)$
and $g_{nc}(r)$ is clearly very much smaller than for the other structure
factors and pair correlation functions, respectively.

Similar data have been taken for two other paths in the phase diagram of
Fig.~\ref{fig2}, namely one at constant temperature, $T=1.5$, varying
$\Delta \mu$ (and hence $x_{\rm B}$), and one at constant $x_B=0.10375$. It
turns out that the radial distribution functions $g_{\alpha \beta}
(r)$ always look similar to those that are shown in Fig.~\ref{fig3}, and
hence are not shown here. The same statement applies to $S_{nn}(q)$, which
exhibits only a weak dependence on either temperature or composition. Here
we hence show only those quantities which have a more pronounced
dependence on our control parameters, namely $S_{cc}(q)$ and $S_{nc}
(q)$, Figs.~(\ref{fig6}),~(\ref{fig7}). From Fig.~\ref{fig6}a we see that
the enhancement of $S_{cc} (q \rightarrow 0)$ persists to rather high
temperatures, and also $S_{nc}(q)$ is only weakly temperature-dependent in
the temperature region shown. In contrast, when we change the composition
at $T=1.5$, we find that $S_{cc}(q)$ quickly loses all structure when
$x_{\rm B} \rightarrow 0$ (Fig.~\ref{fig7}a), and a related trend is
observed in $S_{nc} (q)$.

\section{Dynamic properties of the symmetrical binary fluid mixture}

The detailed description of the structure of our Lennard-Jones
model that we have done so far is necessary to understand its
dynamic properties. The next two sections are devoted to the results
for time-dependent correlation functions and transport coefficients.
In Sec.~5.1 we analyze the dynamic properties of our system along the
fluid-fluid coexistence line whereas in Sec.~5.2 dynamic quantities are
discussed that have been obtained along the path at $x_{\rm B}=0.10375$
which starts in the one-phase region and ends at the coexistence line 
at $T=1.4$ (see Fig.~\ref{fig2}).

\subsection{The dynamics at phase coexistence}

We start by showing the incoherent intermediate structure function
for the A particles, $F_s^{\rm (A)} (q,t)$ \{Eq.~(\ref{eq9})\}, see
Fig.~\ref{fig8} (data for B-particles look very similar and therefore
are not shown). Note that for $q \approx 7$, where the peaks of the
static structure factors $S_{\alpha \beta} (q)$ and $S_{nn}(q)$ occur,
$F_s^{\alpha} (q,t)$, $\alpha \in$ (A,B), decays on the time scale of
about $t=20$ and this decay is certainly even faster for $q=10$. Moreover,
the temperature dependence of the $F_s^{\alpha} (q,t)$ is weak in the
temperature range under consideration.  Such a short structural relaxation
time as well as a weak temperature dependence indeed are expected for any
ordinary fluid, which has no tendency to glass formation \cite{43,44,45}.

On the other hand, for small $q$ the decay of $F_s^{\rm (A)}
(q,t)$, $F_s^{\rm (B)} (q,t)$ is much slower, and this slowing
down simply reflects the diffusive behavior predicted in
Eqs.~(\ref{eq11})-(\ref{eq14}). Fig.~\ref{fig9} shows that this
interpretation in fact is nicely consistent with the data in that,
in agreement with Eq.~(\ref{eq20}), the product $\tau_{\alpha} q^2$
approaches the inverse diffusion constant $D_{\alpha}$
for $q\to 0$ (the values for $D_{\alpha}^{-1}$ are indicated as arrows
in Fig.~\ref{fig9}).  One can further see that the B particles diffuse a
bit faster than the A particles. Of course, we expect that this dynamic
asymmetry must vanish when we are at the critical point, and indeed,
as one can infer from Fig.~\ref{fig9}, the ratio $D_{\rm B}/D_{\rm A}$
decreases to one if the critical point is approached from below along the
coexistence line. The difference in the diffusion constants has a simple
reason: In the A rich phase the B particles find mainly A particles
as nearest neighbors, and since the interaction between particles of
different species is weaker than for equal particles it is easier for
the B particles to escape from an A rich neighborhood than from a B
rich one. Thus, the effective activation energy for the diffusion of B
particles is smaller in an A rich than in an B rich environment.
Note that the behavior of $\tau_{\alpha} q^2$ at finite $q$ is
very different from liquids near a glass transition: In such liquids
one would see a peak in $\tau_{\alpha} q^2$ around a $q$ value that
corresponds to the location of the structure factor maximum (see
Ref.~\cite{60} and references therein).

Particularly interesting is the behavior of the collective structure
functions (Figs.~\ref{fig10},~\ref{fig11}). In $F_{\rm AA} (q,t)$
(Fig.~\ref{fig10}) we observe again a very fast decay for $q$ values
around the structure factor maximum $q=7$ and at $q=0.992$, i.e.~at small
$q$, the decay occurs on a two orders of magnitude larger time scale.  For
the latter value of $q$ one can clearly recognize that several relaxation
processes contribute, since the decay emerges in two steps. Note that
$F_{\rm AB} (q,t)$ and $F_{\rm BB} (q,t)$ exhibit a similar behavior and
so are not shown here. In contrast to the partial structure functions,
in the concentration-concentration correlation function $F_{cc}(q,t)$ only
the slow decay is found at $q=0.992$ (Fig.~\ref{fig11}). Surprisingly, this 
is also the case at $T=1.2$ although, at this temperature, there is almost 
no structure in the corresponding static function $S_{cc}(q)$. As can be
also inferred from Fig.~\ref{fig11} the function $F_{nn}(q,t)$ 
shows a very fast decay to zero at $q=0.992$ (accompanied by oscillations 
due to acoustic modes), whereas it has a slightly slower relaxation to zero than
$F_{cc}(q,t)$ at $q=7$, i.e. the location of the structure factor maximum.

The relaxation times for the collective correlation functions are of
the same order as those detected from the self-correlation functions:
disappointingly, there is no clear indication of critical slowing
down when we approach $T_c$ from below, since in the temperature region
studied the overall Arrhenius-like increase of the relaxation time as $T$
is lowered dominates (the Arrhenius-like behavior is demonstrated below
in the case of the diffusion constants).

\subsection{The dynamics from the one phase region to coexistence}

The qualitative character of the relaxation functions $F_s^{\rm (A)}
(q,t)$, $F_s^{\rm (B)} (q,t)$ and $F_{\rm AA} (q,t)$, $F_{cc} (q,t)$
in the one phase region is rather similar to the behavior found along
the coexistence curve. Therefore, we do not show these functions in any
detail here, but proceed immediately to a counterpart of Fig.~\ref{fig9}
in Fig.~\ref{fig12}, where now the relaxation times $\tau_{\rm A} (q) q^2$
and $\tau_{\rm B}(q) q^2$ extracted from the incoherent intermediate
scattering functions at constant $x_{\rm B}=0.10375$ are shown as a
function of wavenumber $q$. Again one finds that the minority species
diffuses a bit faster than the majority. The reason for the larger 
discrepancies of $D_{\rm \alpha}^{-1}$ and $\tau_{\rm \alpha}(q) q^2$
for $q \to 0$ in the case of the B particles is of course due to the
fact that the statistics is worse for the B particles which is the minority
species.

This difference in the diffusion constants is emphasized in the
Arrhenius plot, Fig.~\ref{fig13}, where the logarithm of $D_{\alpha}$
is plotted vs.~inverse temperature. It is seen that a simple thermally
activated behavior (which would show up as straight lines in this plot)
holds only over very restricted temperature regimes which might be
related to the fact that there are small but significant changes in the
local structure when approaching the coexistence line (see especially
Fig.~\ref{fig4}).  Also plotted in Fig.~\ref{fig13} are the inverse
shear viscosity $\eta^{-1}$ and the interdiffusion constant $D_{\rm
int}$. For an accurate computation of these quantities we have undertaken
a much larger effort than for the single particle quantitities:  At each
temperature we averaged over 50 independent runs of 400000 MD steps to
calculate $\eta$ and $D_{\rm int}$ by means of Eq.~(\ref{eq17}) and
Eq.~(\ref{eq18}), respectively. We show also in Fig.~\ref{fig13} the
interdiffusion constant according to Eq.~(\ref{app_dint}) that is valid
for an ideal mixture. Already for $T<2.5$ $D_{\rm int}$ decouples from the
ideal mixture approximation and thus from the self-diffusion constants,
and $D_{\rm int}$ becomes significantly smaller than the $D_{\alpha}$
if one approaches the coexistence curve.

It is now interesting to check for the Stokes-Einstein relation, which
relates to the self-diffusion constant of a diffusing spherical particle
of diameter $d$ in a fluid of viscosity $\eta$ as
\begin{equation} \label{eq20}
D= \frac{k_BT}{2 \, \pi \eta d }.
\end{equation}
The factor 2 in the denominator of Eq.~(\ref{eq20}) corresponds to the
assumption of slip boundary conditions on the surface of the diffusing
particle (for stick boundary conditions a factor 3 instead of 2 
appears)~\cite{landau}.

Using the data for $\eta$, $D_{\rm A}$, $D_{\rm B}$ and invoking
Eq.~(\ref{eq20}) one obtains the corresponding effective Stokes-Einstein
diameters $d_{\rm A}$, $d_{\rm B}$ shown in Fig.~\ref{fig14}.  The data
for $d_{\alpha}$ show that despite the strong concentration fluctuations
in the system the Stokes-Einstein relation is a good approximation. The
finding that the values for $d_{\rm A}$ and $d_{\rm B}$ are different
is reasonable since the distance for nearest A-B neighbors is slightly
smaller than that for A-A or B-B neighbors and the B particles as the minority
species are mostly surrounded by A particles (see Fig.~\ref{fig4});
thus, the effective hydrodynamic diameter for the B particles is slightly
smaller than that for the A particles. Note that with the assumption
of slip boundary conditions in Eq.~(\ref{eq20}) we obtain reasonable
values around 1 for $d_{\rm A}$ and $d_{\rm B}$ whereas for stick
boundary conditions these values are rather small (around 0.6).

\section{Concluding remarks}

\begin{figure}
\psfig{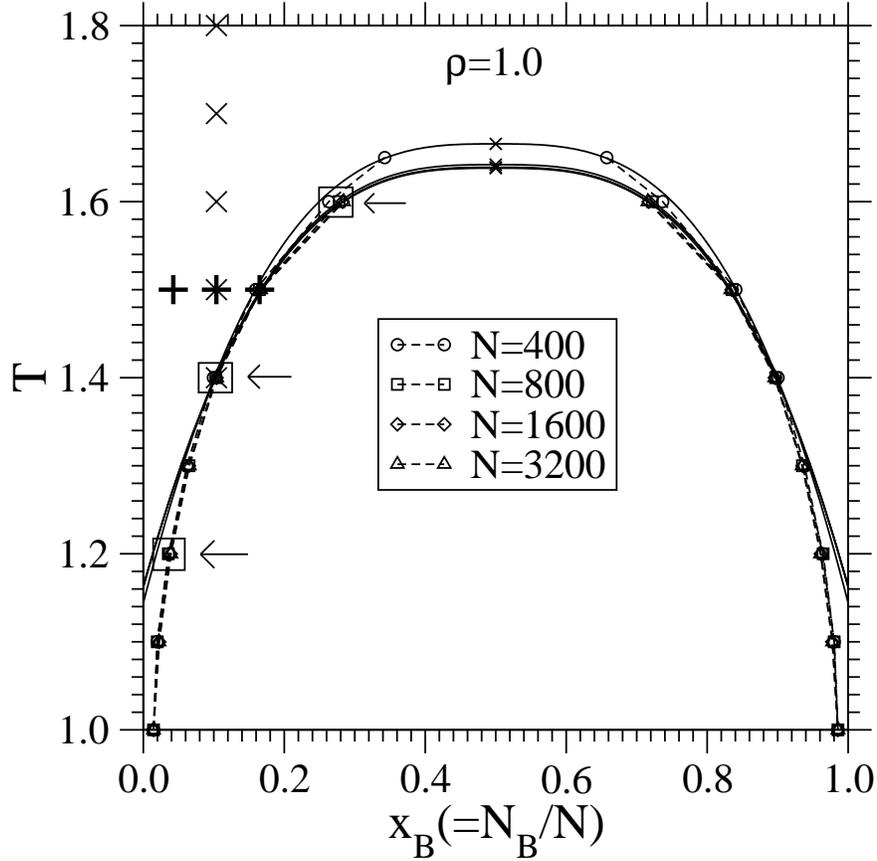}
\caption{\label{fig2}
Phase diagram of the symmetrical binary Lennard-Jones mixture in the plane
of variables temperature $T$ and concentration $x_{\rm B}=N_{\rm B}/N$,
for density $\rho=1.0$ and four choices of $N$, as indicated.  Crosses and
plus symbols indicate paths where the structure and dynamics is studied at
constant concentration $x_{\rm B}=0.10375$ as a function of temperature
and at constant temperature $T=1.5$ as a function of concentration $x_{\rm
B}$, respectively (see Secs.~4 and 5.2). Squares with arrows show states at the
coexistence curve, which will be analyzed in detail in Secs.~4 and 5.1. The
full curves show power law fits with Eq.~(\ref{eq_pow}).  For $N=400$
(thin curve) one obtains $T_c \approx 1.666 \pm 0.005$, while for all
larger values of $N$ the result is $T_c\approx 1.638 \pm 0.005$. The
broken curves are guides to the eye only. The crosses at $x_{\rm B}=1/2$
mark the estimates for $T_c$ that result for the various choices of $N$.}
\end{figure}

\begin{figure}
\vspace*{-3cm}
\psfig{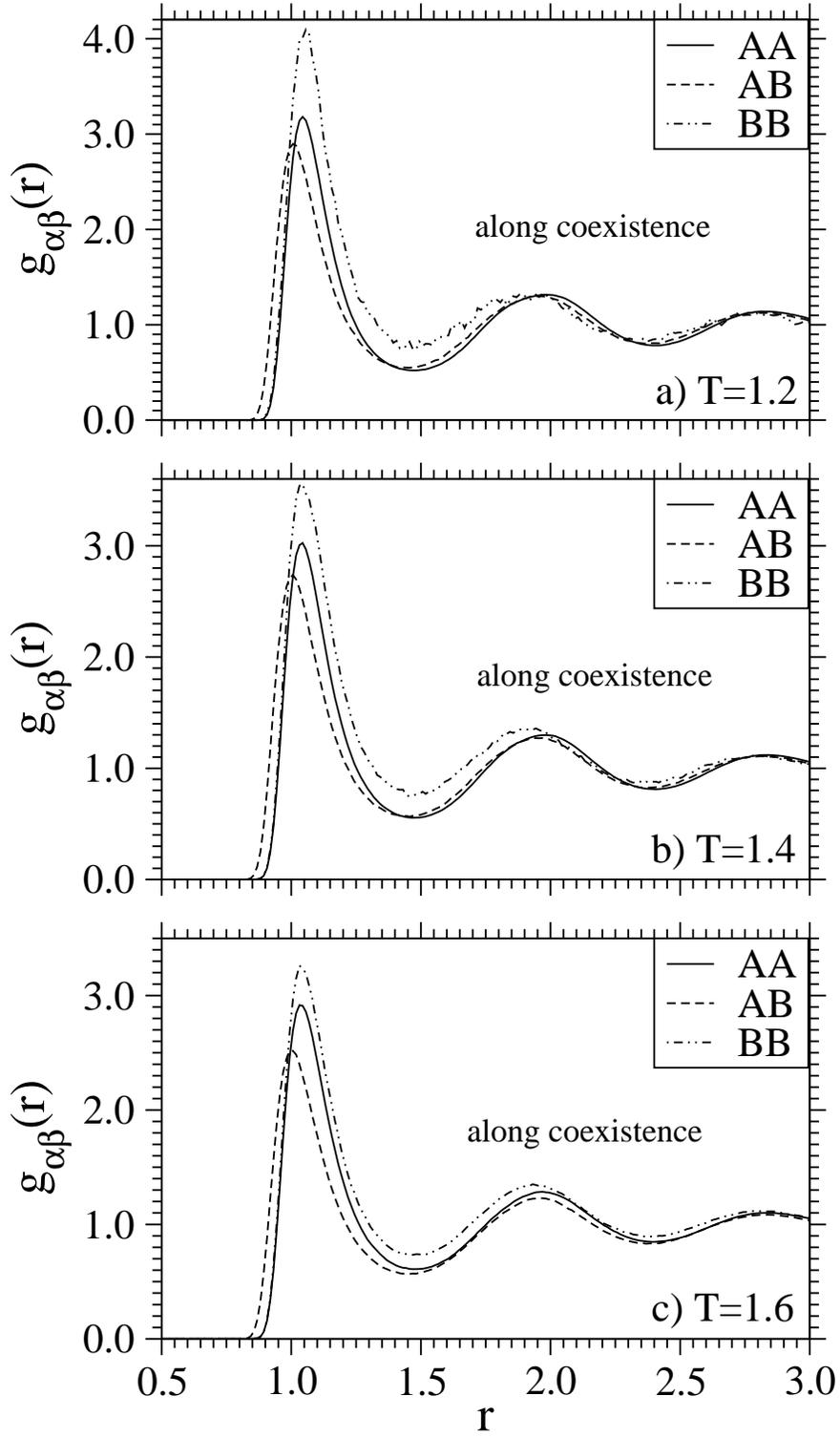}
\caption{\label{fig3}
Radial distribution functions $g_{\rm AA}(r)$ (solid lines), $g_{\rm
AB}(r)$ (dashed lines), and $g_{\rm BB}(r)$ (dashed-dotted lines) along
the A-rich part of the coexistence curve (Fig.~\ref{fig2}) plotted vs.~$r$
at the temperatures a) $T=1.2$, b) $T=1.4$, and c) $T=1.6$.}
\end{figure}

\begin{figure} 
\psfig{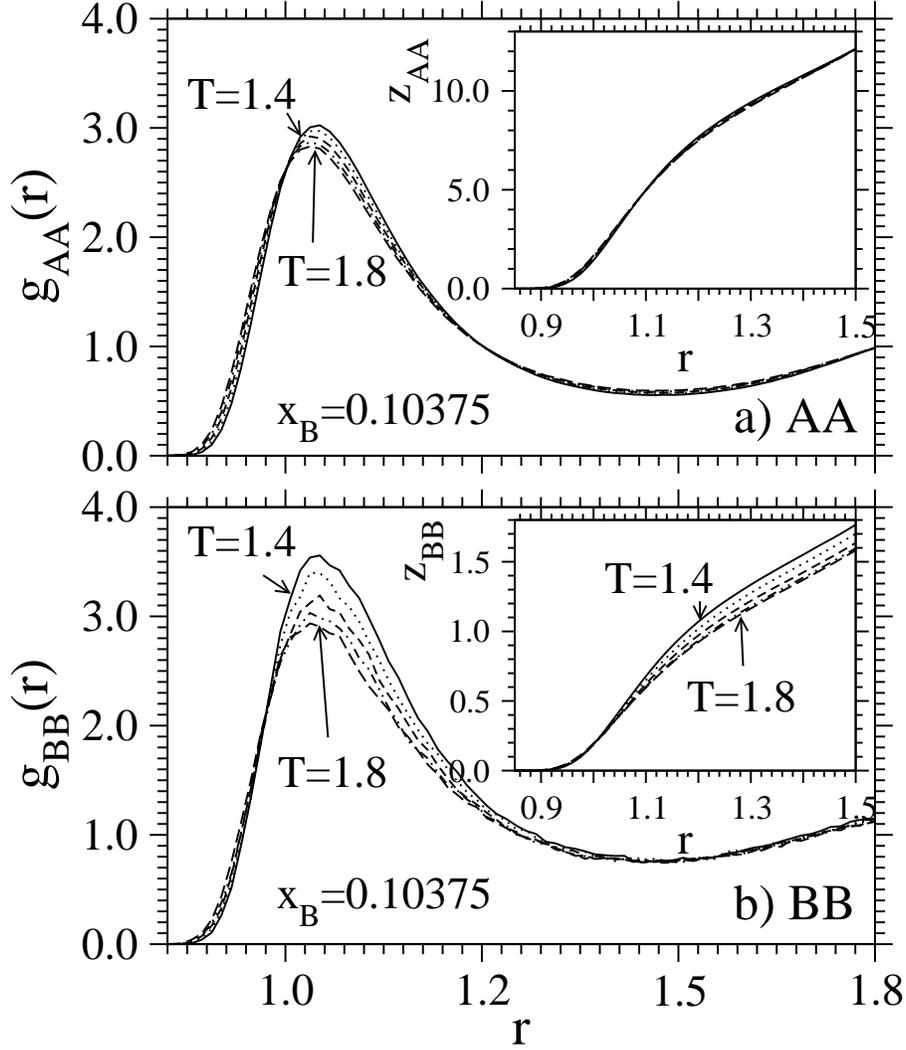}
\caption{\label{fig4}
Radial distribution functions $g_{\rm AA}(r)$, part a), and $g_{\rm
BB}(r)$, part b), for the temperatures $T=1.4$, 1.5, 1.6, 1.7, and 1.8 at
the concentration $x_{\rm B}=0.10375$. The insets show the corresponding
coordination numbers $z_{\alpha \alpha} \quad \alpha \in {\rm [A,B]}$
as a function of $r$.}
\end{figure}

The present paper considers as a simple model system for a binary
fluid mixture with a miscibility gap a dense Lennard-Jones mixture
with fully symmetrical interactions,
$\sigma_{\rm AA}=\sigma_{\rm BB}=\sigma_{\rm AB}=\sigma$,
$\varepsilon_{\rm AA}=\varepsilon_{\rm BB} =\varepsilon=1$,
$\delta=\varepsilon_{AB}-\varepsilon= 0.5$, and also the
masses of the two types of particles are chosen the same,
$m_A=m_B=m$. The aim of the present work was to carry out a
feasibility study, where by combination of the semi-grandcanonical
Monte Carlo technique with Molecular Dynamics simulations both
static properties, including the phase diagram, and dynamic
correlation functions and associated transport coefficients are
obtained simultaneously. It is shown that outside of the critical
region of the mixture rather small system sizes (such as a total
number of $N=1600$ particles) already suffice to obtain
quantitatively reliable results, while in the critical region
(which could not yet be studied here) much larger sizes clearly
are mandatory. Such large sizes would also be needed for the very
interesting problem of spinodal decomposition of the binary fluid
mixture after a quench from the one-phase region into the unstable
region of the phase diagram (Fig.~\ref{fig2}). Since our study is
the first study of a fluid mixture where both the phase diagram
and the relevant transport coefficients have been determined
simultaneously, such an extension to far from equilibrium dynamics
clearly would be very interesting.

Our studies of transport coefficients are distinct from earlier
work on similar models by the fact the we know precisely where in the
phase diagram the considered state points are for which transport
coefficients have been determined. This fact clearly facilitates
the proper interpretation of the results. A particular interesting
finding concerns the asymmetric composition in our otherwise fully
symmetric model.

In further work we also plan to study systematically Onsager
coefficients to test the concepts of phenomenological irreversible
thermodynamics for the present model and to check ``mixing rules'' for
the interdiffusion coefficient. Moreover, we will present extensions to more
realistic models of real materials which always exhibit asymmetry
in the interactions, to allow also a comparison to suitable
experiments.

\bigskip

{\bf \underline{Acknowledgments}:} The present research was
supported by the Deutsche Forschungsgemeinschaft (DFG) under grant
N$^o$ Bi314/18 (SPP 1120). One of the authors (J. H.) acknowledges
the support of the DFG under grant N$^o$ HO 2231/2-1.

\clearpage

\begin{figure}
\hspace*{-1.3cm}
\psfig{file=./fig5a.eps,height=12cm}
\end{figure}

\begin{figure}
\vspace*{-4cm}
\hspace*{0.7cm}
\psfig{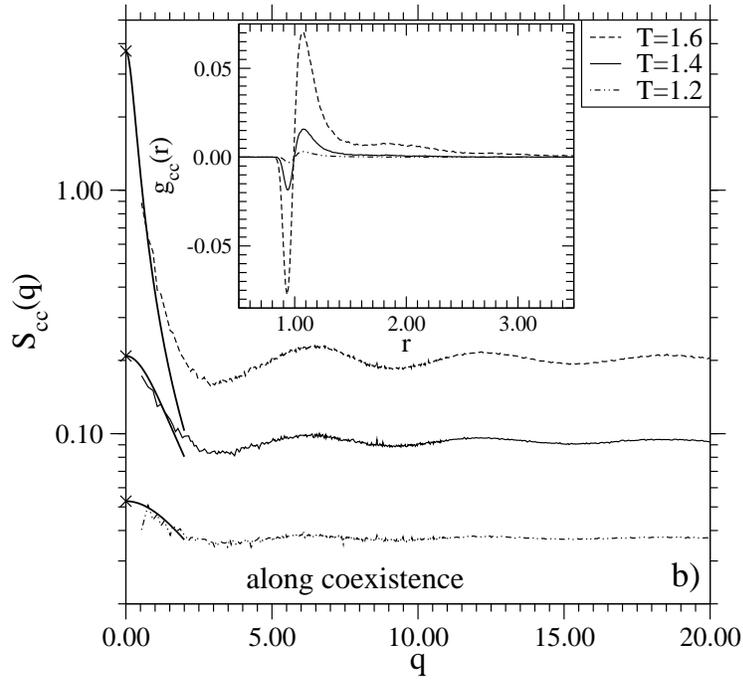}

\vspace*{1cm}
\hspace*{0.7cm}
\psfig{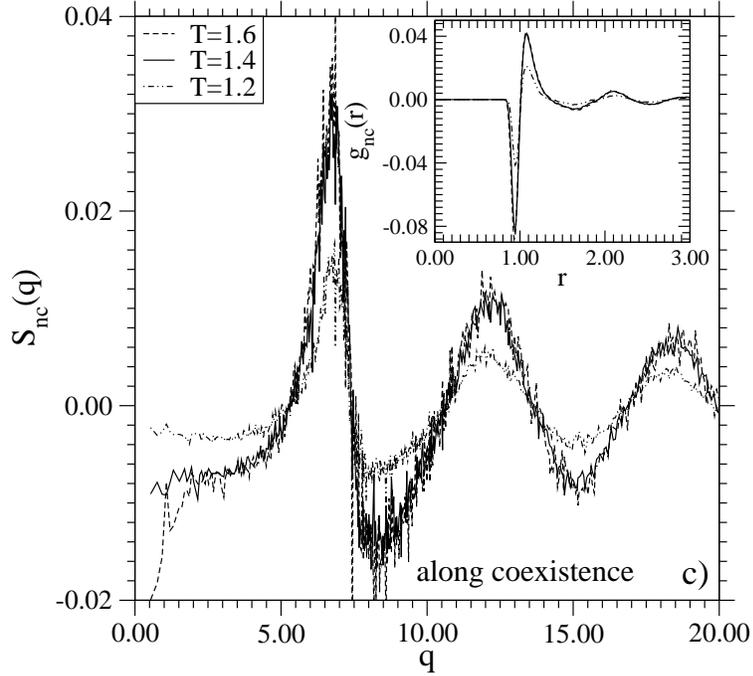}
\caption{\label{fig5}
Different structure factors plotted vs.~$q$ for the three indicated
temperatures along the coexistence curve. a) $S_{nn} (q)$, b) $S_{cc}(q)$
(the bold solid lines are fits with Eq.~\ref{eq8a}, the crosses are
estimated of $S_{cc}(q=0)$ as obtained from Eqs.~(\ref{eq3}) and
(\ref{eq8a})), and c) $S_{nc}(q)$. Note that a logarithmic scale is
used on the ordinate of part b). The insets in b) and c) show the
corresponding functions in real space.}
\end{figure}

\begin{figure}
\vspace*{-4cm}
\hspace*{0.7cm}
\psfig{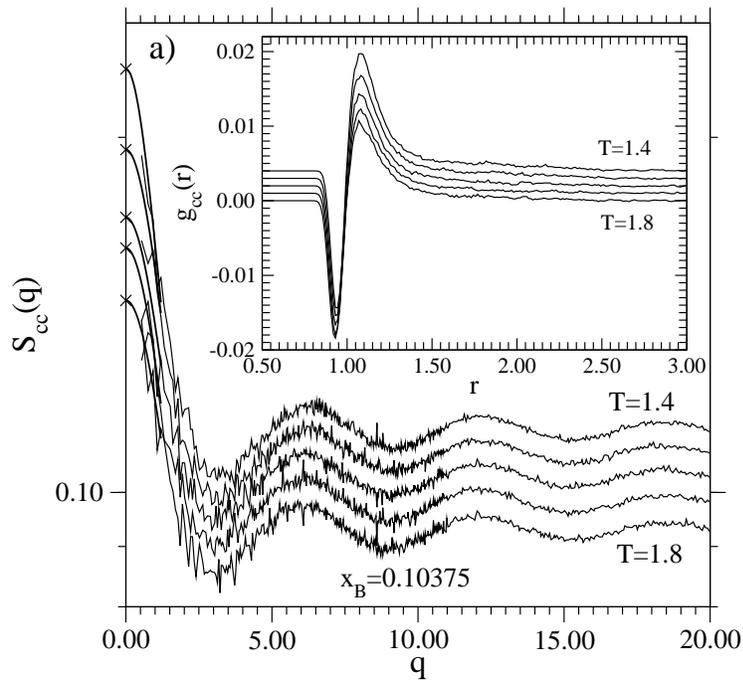}

\vspace*{1cm}
\hspace*{0.7cm}
\psfig{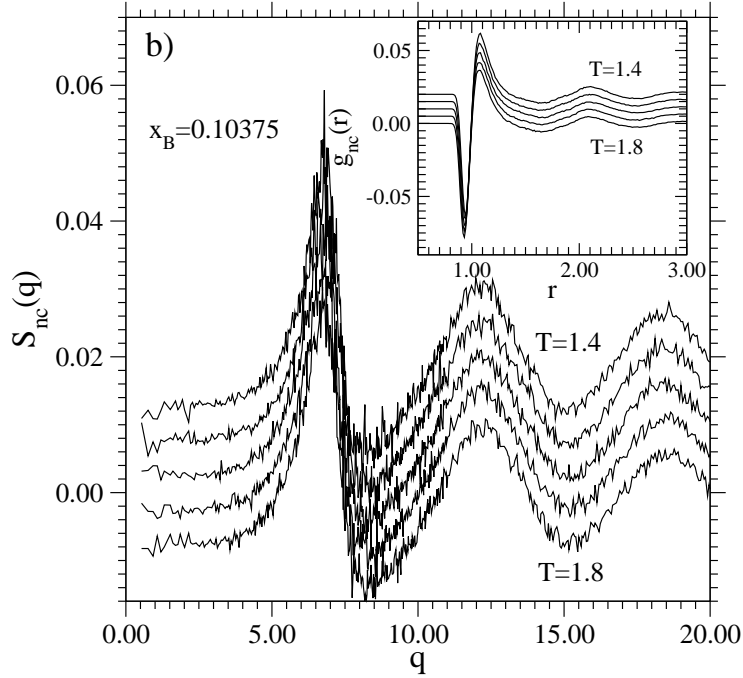}
\caption{\label{fig6}
Structure factors $S_{cc}(q)$, part a), and $S_{nc} (q)$, part b), plotted
vs. $q$, for the temperatures $T=1.4$, 1.5, 1.6, 1.7, and 1.8 (from top
to bottom), at the concentration $x_{\rm B}=0.10375$. For $S_{cc}$ again
a logarithmic scale is used. The curves are overshifted relative to each
other by an amount $\Delta = 0.005$ ($T=1.8$ is the unshifted curve). The
crosses in part a) are estimates of $S_{cc}(q=0)$ as in Fig.~\ref{fig5}b 
and the bold solid lines are fits with Eq.~\ref{eq8a}. The insets
show the corresponding functions in real space (also overshifted to
each other by an amount $\Delta=0.001$ in part a) and $\Delta=0.005$
in part b), again $T=1.8$ is the unshifted temperature).}
\end{figure}

\begin{figure}
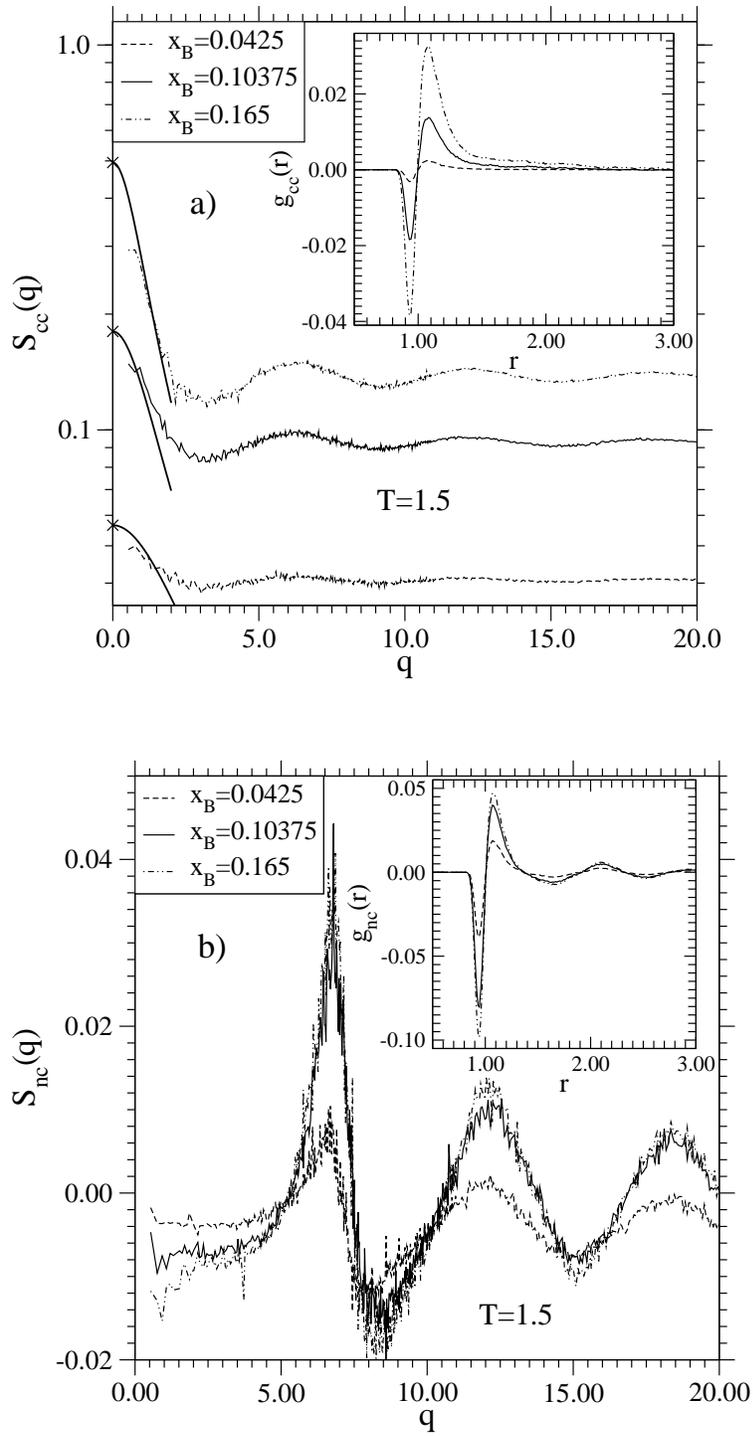

\vspace*{-3cm}
\hspace*{0.7cm}
\psfig{file=./fig7a.eps,height=9cm}

\vspace*{1cm}
\hspace*{0.7cm}
\psfig{file=./fig7b.eps,height=9cm}
\caption{\label{fig7} 
Same as Fig.~\ref{fig6}, but at constant temperature $T=1.5$ and three 
concentrations as shown (note that all the curves are unshifted).}
\end{figure}

\begin{figure}
\psfig{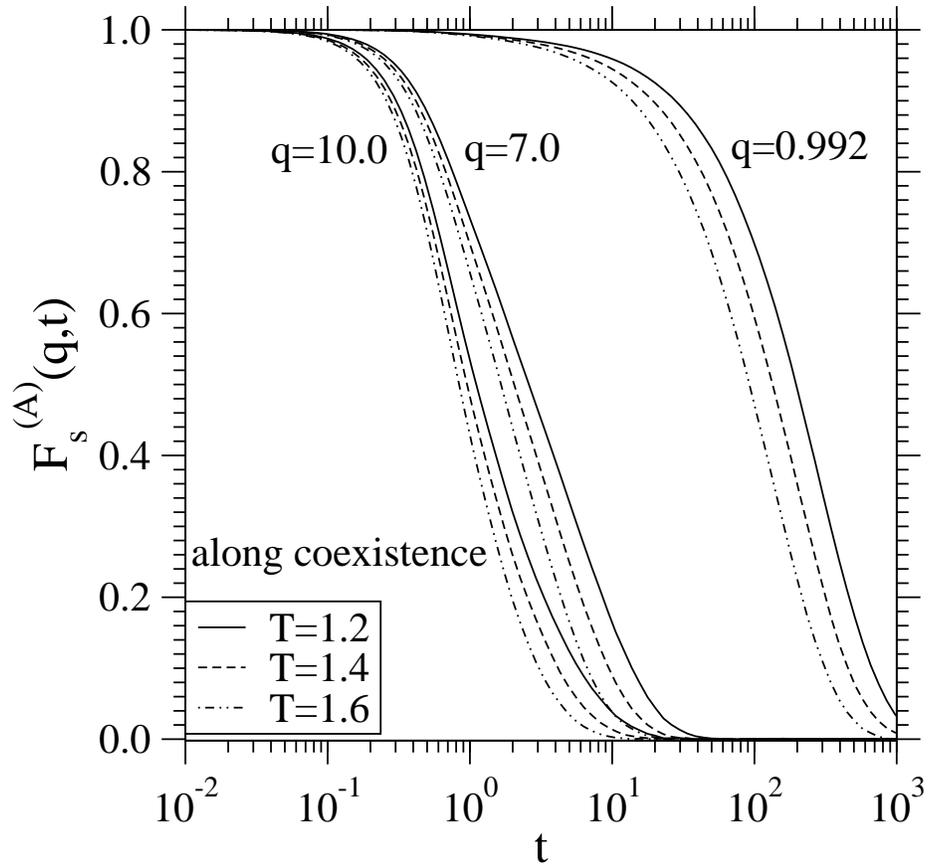}
\caption{\label{fig8}
Incoherent intermediate scattering function of A particles plotted versus
time (note the logarithmic scale of time) for the three temperatures
$T=1.2$ (solid lines), $T=1.4$ (dashed lines), and $T=1.6$ (dashed-dotted
lines) along the coexistence curve for three different values of $q$,
namely: $q=0.992$, 7, 10 (from right to left).}
\end{figure}

\begin{figure}
\psfig{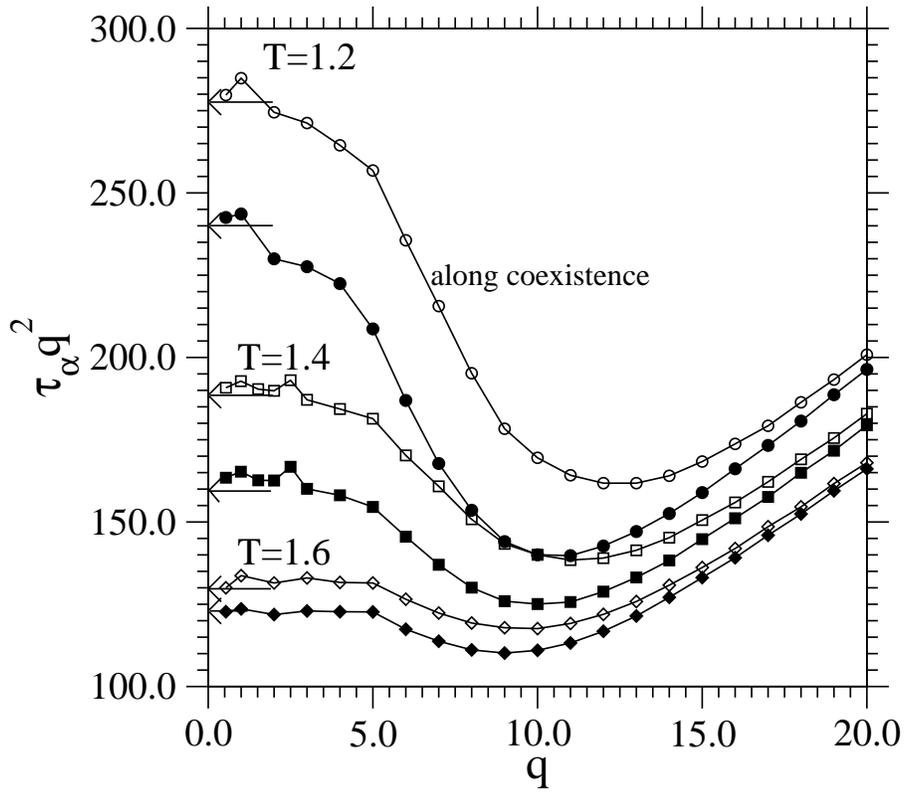}
\caption{\label{fig9}
Plot of scaled relaxation time $\tau(q) q^2$ vs. $q$, for the A-particles
(open symbols) and the B-particles (filled symbols) for three temperatures
along the A-rich branch of the coexistence curve. The arrows indicate
the value of the corresponding inverse self-diffusion constants.}
\end{figure}

\begin{figure}
\vspace*{-3cm}
\psfig{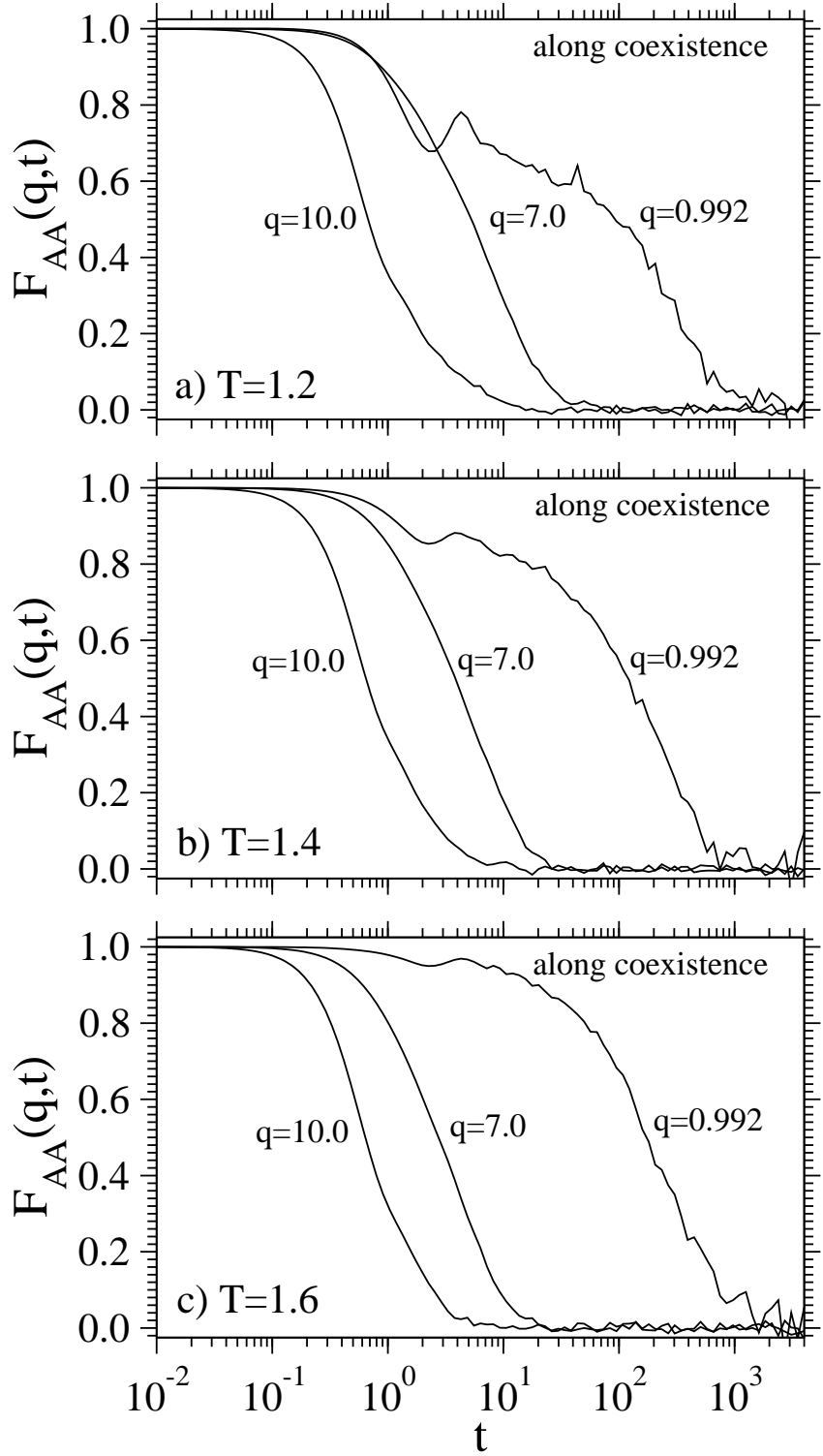}
\caption{\label{fig10}
Coherent intermediate scattering function
$F_{\rm AA}(q,t)$ of A-particles plotted vs.~time (note logarithmic
scale of time) for three temperatures and concentrations given
from the A-rich part of the coexistence curve. Three temperatures
are shown, $T=1.2$, part a), $T=1.4$, part b), and
$T=1.6$, part c). Three wavenumbers $q$ are included as
indicated.}
\end{figure}

\begin{figure}
\vspace*{-3cm}
\psfig{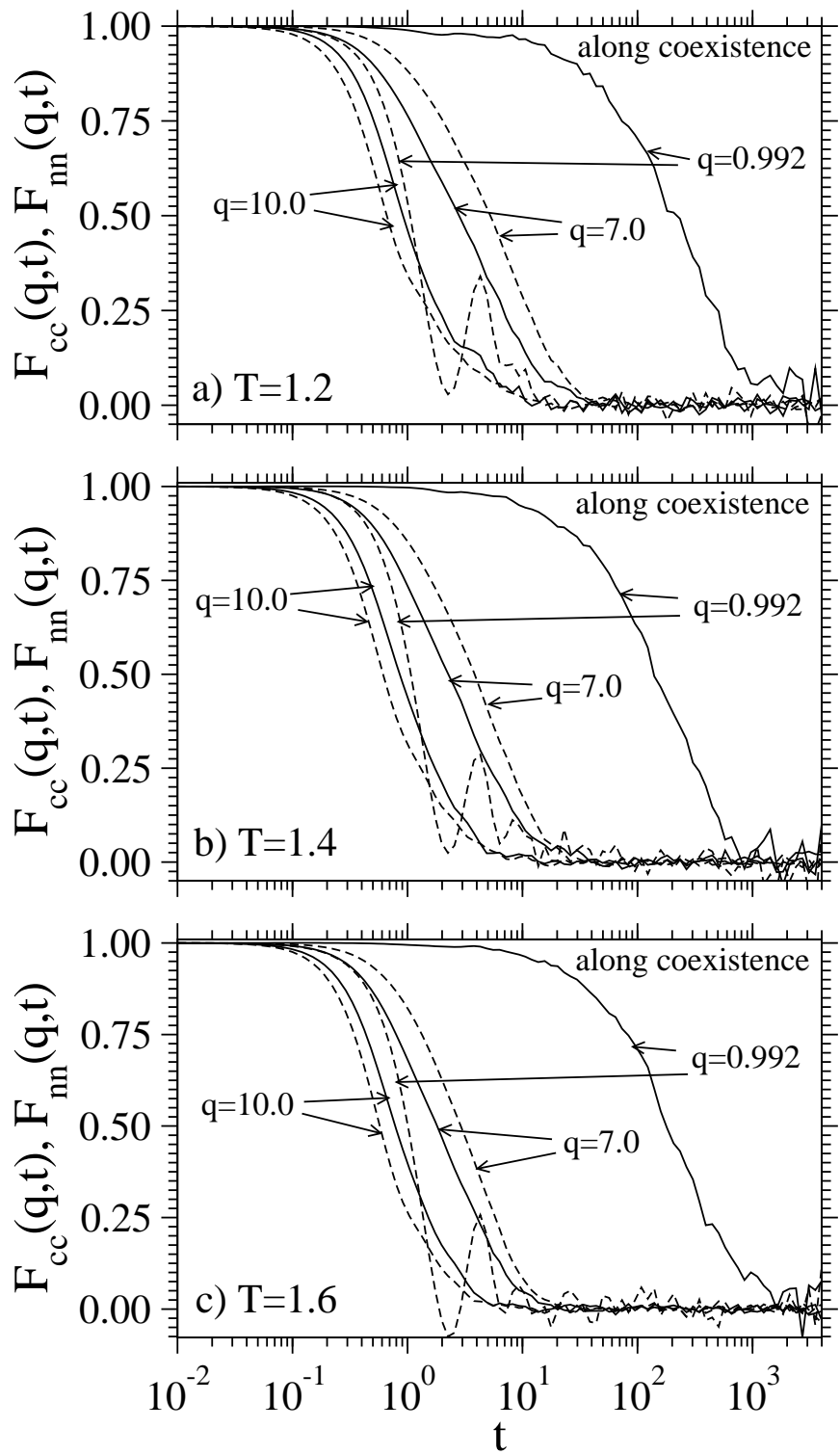}
\caption{\label{fig11}
Same as Fig.~\ref{fig10} but for $F_{cc}(q,t)$ (solid lines) and
$F_{nn}(q,t)$ (dashed lines).}
\end{figure}

\begin{figure}
\psfig{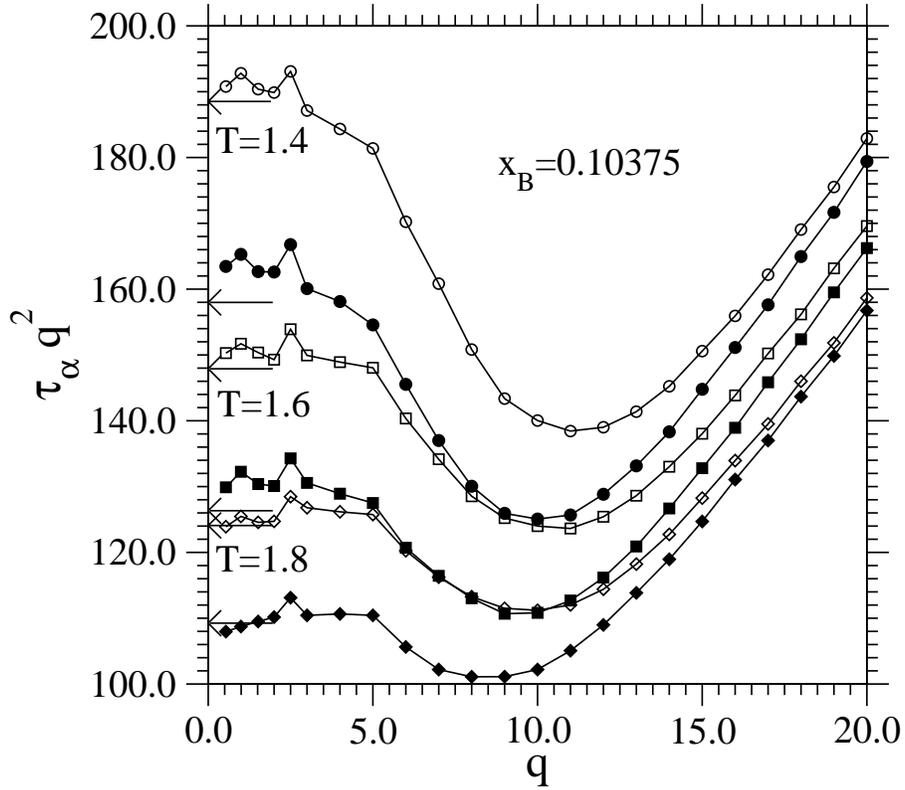}
\caption{\label{fig12}
Scaled relaxation times $\tau_{\rm A}(q) q^2$ for the A-particles
(open symbols) and $\tau_{\rm B}(q) q^2$ for the B-particles (filled
symbols) plotted vs.~wavenumber $q$. Three temperatures at constant
$x_{\rm B}=0.10375$ are shown. The arrows indicate the values of the
corresponding inverse diffusion constants, as extracted from the mean
square displacements via the Einstein relation.}
\end{figure}

\begin{figure}
\psfig{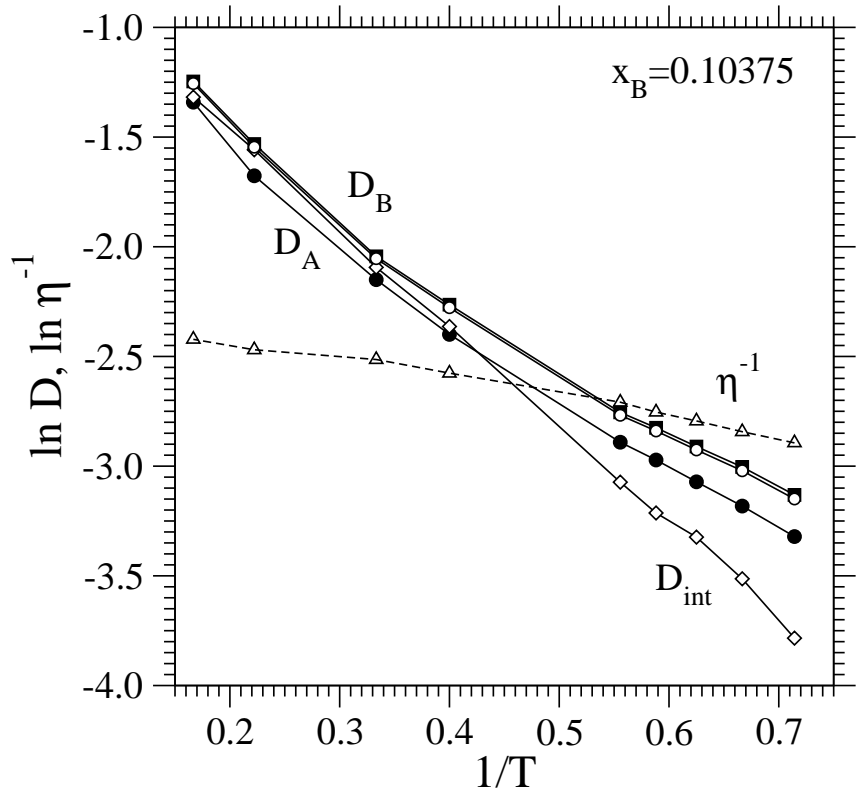}
\caption{\label{fig13}
Logarithm of the self-diffusion constants (filled symbols)
and the interdiffusion constant (open diamonds) as well as of the inverse
viscosity (open triangles) plotted vs.~$1/T$ for a wide range of temperatures for
the mixture at constant concentration $x_{\rm B}=0.10375$. Also plotted is 
$D_{\rm int}$ as calculated from Eq.~(\ref{app_dint}) (open circles). The viscosity
is multiplied by a factor of 3. The lines are guides to the eye.}
\end{figure}

\begin{figure}
\psfig{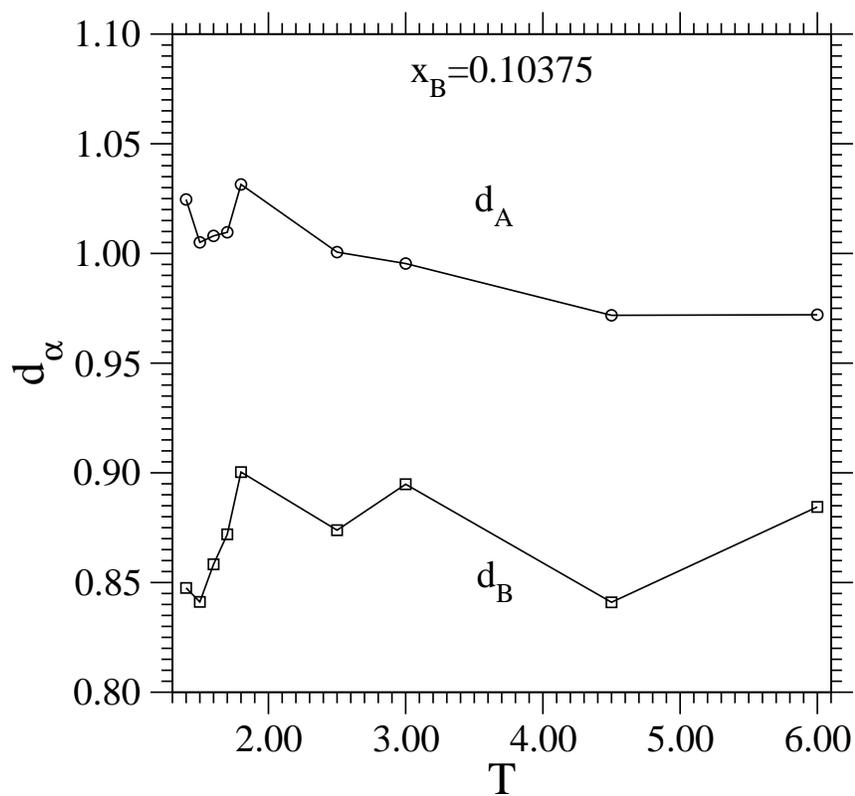}
\caption{\label{fig14}
Stokes-Einstein diameters $d_{\rm A}$, $d_{\rm B}$ of
the diffusing particles plotted vs.~temperature, for the mixture
at constant concentration $x_{\rm B}=0.10375$.}
\end{figure}

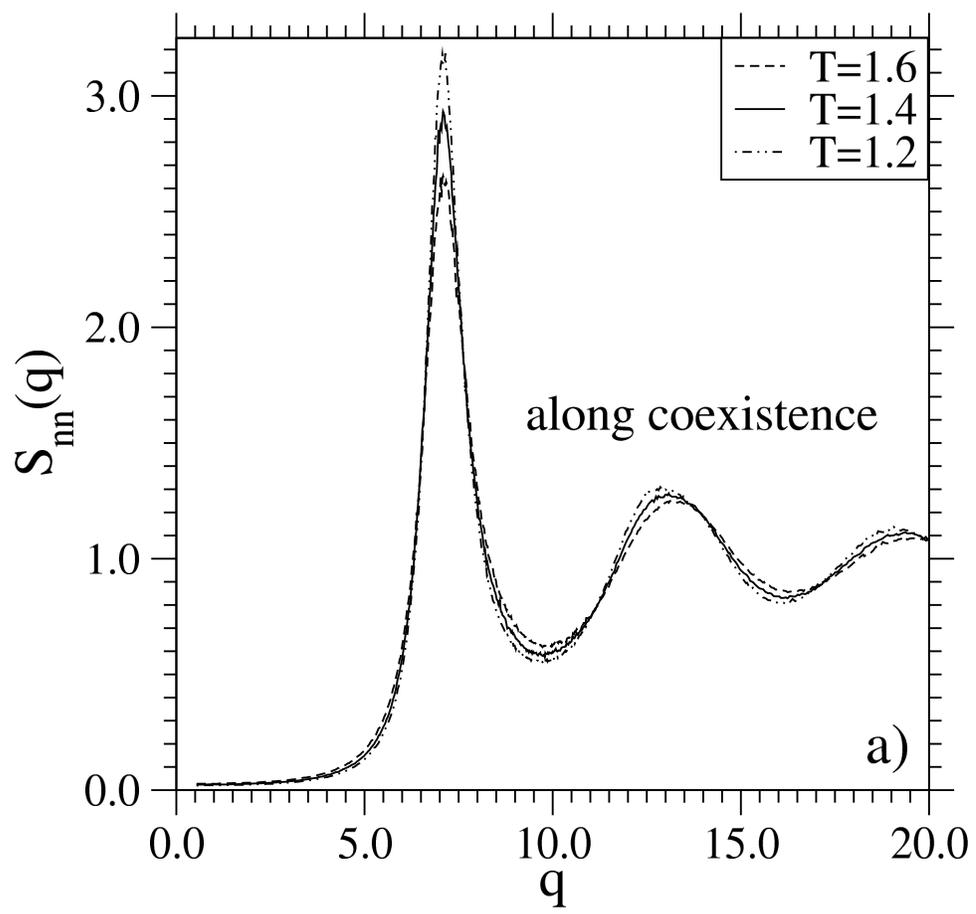
\end{document}